*Review*

# Reconfigurable Intelligent Surfaces for 6G: Applications, Challenges and Solutions

Yajun Zhao[1,2,*], IEEE Member, Xin Lv[1]

[1] Beijing Institute of Technology, Beijing 100081, China;
[2] ZTE Corporation, Beijing 100029, China;
[*] Correspondence: zhao.yajun1@zte.com.cn; Tel.: +86 13910120374

**Abstract:** It is expected that scholars will continuously strengthen the depth and breadth of theoretical research on RIS, and provide a higher theoretical upper bound for the engineering application of RIS. While making breakthroughs in academic research, it has also made rapid progress in engineering application research and industrialization promotion. This paper will provide an overview of RIS engineering applications, and make a systematic and in-depth analysis of the challenges and candidate solutions of RIS engineering applications. Future trends and challenges are also provided.

## 1. Introduction

6G is considered to be used to meet the needs of wireless communication for more than ten years after 2030, to support the intelligent society in the future [1]. To realize the vision of 6G, many potential key technologies have been proposed, including artificial intelligence, MIMO, RIS, and so on [2]. Among them, RIS has received great attention from academic and industrial circles due to its two-dimensional super surface structure, intelligent programmability, and the ability to abnormally regulate electromagnetic waves, as well as its technical characteristics of low cost, low power consumption, and simple deployment.

At present, from the academic (theoretical) point of view, the research achievements of RIS have laid a solid foundation for engineering applications. It is expected that scholars will continuously strengthen the depth and breadth of theoretical research on RIS, and provide a higher theoretical upper bound for the engineering application of RIS. While making breakthroughs in academic research, it has also made rapid progress in engineering application research and industrialization promotion. As early as 2018, NTT DoCoMo was the first to trail on RIS [3]. However, in the following two years, RIS was still mainly active in academic research field, and did not receive wide attention in the industrial field. In June 2020, ZTE Corp. together with Southeast University and other organizations, promoted and established the "RIS Task Force" in China IMT-2030 (6G) Promotion Group, which absorbed more than 30 universities and enterprises to jointly promote the technical research, standardization and industrialization of RIS. The establishment of the task force made RIS begin to attract wide attention from the industry, and really promoted RIS from academia to industry [4]. Since 2020, domestic academic and industrial circles have jointly carried out a series of RIS industry promotion activities, which greatly promoted the technological research and industrialization of RIS. In September 2020, ZTE Corp. joined hands with more than ten domestic and foreign enterprises and universities such as China Unicom to establish the "RIS Research Project" in CCS A TC 5-WG 6. On September 17th, 2021, IMT-2030 (6G) Promotion Group officially released the industry's first research report on intelligent hypersurface technology in the 6G seminar. On September 24th, 2021, ZTE Corp., Southeast University, China Unicom, etc. Jointly hosted the "1st Reconfigurable Intelligent Surface Technology Forum". On April 7th, 2022, RIS Technology Alliance (RISTA) was established and the first general meeting of RISTA members was held in Beijing[1].

With the in-depth research of RIS application, the challenges faced by RIS have gradually been exposed. If these challenges cannot be solved well, RIS will be difficult to be deployed on a large scale. Problems faced by RIS in engineering applications have been analyzed and studied, but the number of literature and related achievements are still insufficient. Based on our previous work [5][6], this paper will make a further systematic and in-depth analysis of the challenges and candidate solutions of RIS engineering applications. According to our research, although the engineering application of RIS will face many challenges, there are corresponding solutions to overcome these challenges. This paper will focus on the typical problems in the actual network deployment of RIS, including multi-user access, network coexistence, and so on, hoping to sort out the existing work of the system and promote research in this area.

---

[1] http://www.risalliance.com/LiveVideoServer/risWeb/events_202204_en.html

This article structure is as follows. The engineering application of RIS is summarized in the second part. In the third part, the technical challenges and solutions of RIS engineering applications will be discussed. The fourth part will give a brief introduction to the future trends and challenges. Finally, conclusions are drawn.

**2. Overview on Engineering Applications of RIS**

This section will first briefly discuss the evolution from traditional multiple antenna technology to RIS. Then the engineering application of RIS will be outlined, including RIS technical characteristics, implementation types and typical network deployment scenarios.

*3.1. Evolution from Multiple Antennas to RIS*

This part will discuss three technical concepts, including the enhancement of MIMO technology, the basic concepts, and the development of massive MIMO cell-free and RIS. We try to explore the logical relationship between these three technological evolutions from two angles.

3.1.1. Utilization and transformation of wireless channel

It does not change the propagation environment of the natural wireless channel, that is, passively adapts to natural wireless channels. To make full use of the wireless channel and approach the capacity upper bound of the natural channel, a traditional and classic way is to continuously increase the number of MIMO antennas, to fully obtain the benefits of array gain and spatial multiplexing. However, as the number of antennas increases, the performance improvement brought by them is close to the ceiling.

Another direction of MIMO technology evolution is from centralization to distribution. From centralized MIMO to distributed MIMO (D-MIMO/CoMP), the wireless channel is segmented by distributed antennas. The channel transformation of each segment is reduced, and the change rate and variance of the channel are reduced (obtaining macro diversity; taking channel segmentation quantization, segmentation regulation, and cooperative regulation). If the distributed antenna density is increased and the channel is further segmented, the change of each channel segment in time can be ignored, and the entire wireless channel becomes an approximate constant parameter channel with a known channel in time. However, the requirements for deployment site, synchronization, power supply and return link brought by distributed antennas, as well as the high-cost problems caused by them, will make it difficult to implement densely deployed distributed antennas in engineering. Due to the complexity of its engineering implementation, the CoMP features, which have been proposed and standardized in the 4G stage, have not been well realized in the network so far.

Whether it is to increase the antenna size or to deploy in a distributed and cooperative way, it is to reach the upper limit of the capacity of the natural wireless channel as much as possible without changing the propagation environment of the natural wireless channel, that is, it belongs to the optimization mechanism of the passive adaptive channel. If we can actively change the natural propagation channel, it may bring a new possibility to improve the performance of wireless communication systems, and the emergence of RIS makes it possible. RIS has the characteristics of passive regulation, low cost, and easy deployment, which may be widely deployed in the natural propagation environment, allowing it to truly regulate the wireless channel.

3.1.2. From the perspective of network architecture evolution: break the cellular limit and bring a new network paradigm

From the perspective of the evolution of network architecture, the evolution of MIMO technology has changed from centralized MIMO to distributed MIMO (D-MIMO/CoMP), thus breaking the limitation of cellular to cell-free. From the perspective of optimization, the basic idea is to optimize the transceiver nodes, that is, from the point-to-point or point-to-multi-point of the transceiver antenna set in the cell to break the cell constraints, to achieve multi-point to multi-point joint optimization of a larger transceiver antenna set. It can be seen that the network architecture evolution of traditional MIMO technology is limited to the optimization of the distribution of transceiver nodes and the expansion of cooperation sets, and the wireless channel itself is not considered as an component of the network architecture evolution.

The ubiquitous deployment of RIS makes the wireless channel a part of the network. By controlling the wireless propagation environment through RIS, the joint optimization of transceiver nodes and wireless channels is realized, bringing a new paradigm to the evolution of the future network architecture.

*3.2. Technical Features and Typical Classification of RIS*

The main technical features of RIS can be summarized as follows (refer to Figure 1).

**A reconfigurable artificial two-dimensional meta-surface:** From three-dimensional meta-materials to two-dimensional meta-surfaces, the interface effect produces amplitude and phase abrupt changes, which are different from the integral effect of three-dimensional meta-materials.

**Passivity regulation:** In addition to the controller, RIS does not require a dedicated power supply to regulate radio waves.

**Abnormal regulation:** RIS can abnormally regulate the propagation characteristics of radio waves such as phase, polarization mode, amplitude and delay. In particular, RIS can negatively refract and reflect electromagnetic waves.

**Channel cascading:** RIS assisted communication channel is the combination of BS-RIS and RIS-UE segmented channels, which requires new channel estimation and beam shaping design, bringing new challenges to algorithm design.

**AI Driven:** AI enabled channel estimation, beam shaping and resource management for RIS to achieve a smart wireless environment.

RIS technology is significantly different from traditional technologies, such as massive MIMO [7], CoMP [8], MIMO relay [9], backscatter [10], and smart repeater [11] (refer to Table 1). From this comparison, we can see that although these traditions may be better than RIS in some technical indicators, they do not have all the advantages of low power consumption, low cost, and easy deployment like RIS. Obviously, only RIS has the potential to support green communication, ubiquitous deployment, and wireless environment reconstruction.

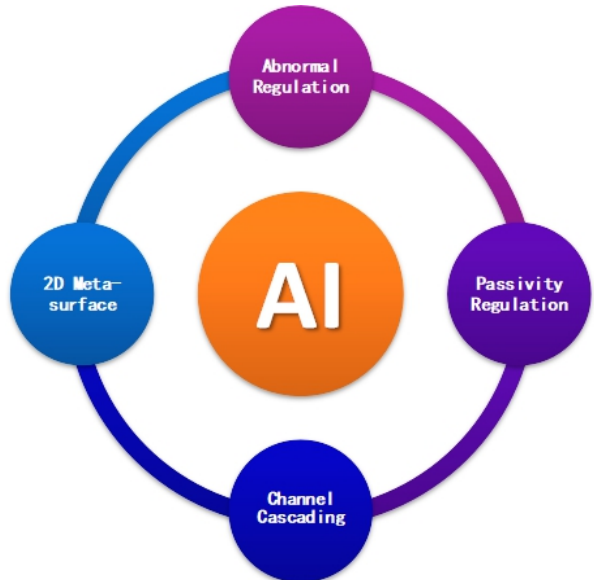

**Figure 1.** Technical Features of RIS.

Table 1. Comparison between RIS and traditional technologies

| Technologies | Operating mechanism | Duplex | RF chains | Thermal noise | Hardware cost | Energy consumption | Role | Architecture |
|---|---|---|---|---|---|---|---|---|
| RIS | Passive/[Active]* Tune | Full duplex | No | No | Low | Low | Helper | Distributed |
| **Backscatter** | Passive<br>Tune | Full duplex | No | No | Very low | Very low | Source | Distributed |
| **MIMO Relay** | Active<br>Receive and transmit | Half/[full] duplex | Yes | Yes | High | High | Helper | Distributed |
| **Massive MIMO** | Active<br>Transmit and receive | Half/[full] duplex | Yes | Yes | Very High | Very high | Source/<br>Destination | **Centralized** |
| **CoMP/D-MIMO** | Active<br>Transmit and receive | Half/[full] duplex | Yes | Yes | Very High | Very high | Source/<br>Destination | Distributed |
| **Smart-repeater** | Active/**[Passive?]**<br>Receive and transmit | Half/[full] duplex | Yes | Yes | Medium | Medium | Helper | Distributed |

*\* Literature [11] also gives an active RIS with low rated power. This type of RIS adopts simple low-power amplifier transistor circuit, so it has the characteristics of low cost and low power consumption.*

As can be seen from the above technical features, the main technical advantages of RIS are supporting abnormal regulation, low power consumption, low cost, simple structure, and easy deployment. These technical advantages make it possible for RIS to be widely deployed in wireless networks, thus making it possible to build an intelligent and controllable wireless environment and bring a new paradigm for future wireless networks (refer to Figure 2).

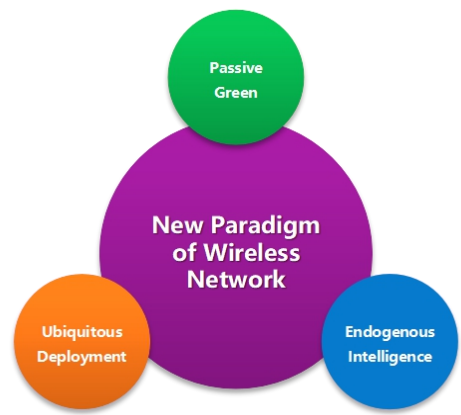

**Figure 2.** Bring a new paradigm for future wireless networks.

With the advancement of RIS engineering research, a variety of different technical characteristics and forms of RIS have appeared in the industry. Here, an attempt is made to classify and summarize the types of RIS reported in the open literature (see Table 2). Each of these types of RIS has its technical characteristics. What kind of RIS can be deployed on a large scale is uncertain till now.

Table 2. RIS types

| **Features** | **Types** |
|---|---|
| **Refraction/Reflection** | Refractive RIS |
| | Reflective RIS |
| | Simultaneously Transmitting and Reflecting RIS (STAR-RIS) [13] |
| **Regulation functions** | Spatial modulation, RIS based transmitter [14][15] |
| | Radio channel regulation, Intelligent reconfiguration of wireless channels |
| | RIS-based novel phased array antenna |
| | RIS-based simultaneous wireless information and power transfer |
| **Regulation mechanisms** | PIN diode tube, varactor diode, MEMS, liquid crystal, Graphene, etc. |
| **Frequency band** | Sub-6GHz |
| | Millimeter-Wave Band |
| | Terahertz band |
| | Optical band |
| **Active or passive** | Passive RIS |
| | Active RIS [16] |
| **Dynamicity** | Static state |
| | Semi static regulation |
| | Dynamic regulation |
| **Measurement/Sensing** | Consists of passive elements only |
| | Including passive elements and some active elements with measuring/sensing ability [17] |
| **Deployment Mode** | Network-controlled |
| | Standalone [18][19] |

*3.3. Typical Deployment Scenarios of RIS*

As can be seen from the technical characteristics of RIS mentioned above, RIS has the characteristics of low cost, low power consumption and simple deployment, and can be widely deployed in future wireless networks. In the previous article, we discussed some typical RIS deployment scenarios. [6][20] Here, we try to classify typical network deployments of RIS according to the characteristics of application scenarios.

This section will classify and summarize the typical deployment scenarios of RIS from four aspects: development mode, coexistence and sharing, increasing rank and coverage, and coverage area.

From the perspective of network deployment mode, RIS deployment scenarios can include standalone mode and network control mode. The two modes are different in requirements of control link, the interaction of measurement/control signaling and the complexity of network deployment, and each has its advantages and disadvantages. For a detailed discussion, please refer to section 3.3.

From the perspective of coexistence and sharing, RIS deployment scenarios can include multiple operator networks coexistence, single-user access and multi-user access, multiple RISs deployment, and spectrum properties (such as licensed spectrum, and unlicensed spectrum).

From the perspective of rank increasing and covering enhancement, RIS deployment scenarios can include deployed near NB, deployed at the cell edge, deployed in the middle of the cell, and ubiquitous deployment.

From the perspective of rank increasing and covering enhancement, RIS deployment scenarios can include remote areas, urban indoor/outdoor, NTN, and so on.

Table 3 provides the four types of RIS deployment scenarios, including typical deployment forms and their respective key points.

Table 3. Four types of RIS deployment scenario

| Deployment Scenarios | Forms | key Points |
|---|---|---|

| **Deployment Mode** | (1) Network Controlled Mode<br>(2) Standalone Mode | ① Control link<br>② Measurement and control signaling |
|---|---|---|
| **Coexistence and sharing** | (1) multiple operator networks coexistence.<br>(2) Single user access and multi-user access.<br>(3) Multiple RISs deployment.<br>(4) Spectrum properties, such as licensed spectrum or unlicensed spectrum. | ① Identification and collaboration of multiple RIS<br>② Interference and coordination of coexistence and sharing in networks |
| **Increasing Rank and Coverage** | (1) Deployed near NB<br>(2) Deployed at the cell edge<br>(3) Deployed in the middle of the cell.<br>(4) Ubiquitous deployment. | ① Increasing Rank<br>② Covering Enhancement<br>③ Near field[Note1]/far field |
| **Coverage Area** | (1) Remote area<br>(2) Urban indoor/outdoor<br>(3) NTN, UAV | ① Wide area: semi-dynamic/static adjustment, which is used to improve capacity and coverage.<br>② Local: Dynamic and accurate channel regulation.<br>③ Challenges of Power Supply and Control Link Deployment. |

Note 1:
① RIS is widely deployed, and near field becomes a typical scenario.;
② RIS passive regulation and low radiation, suitable for near-field communication;
③ The near-field of active phased array antenna may only be applicable to IoT.

## 3. Technical Challenges and Solutions

To realize the above engineering applications, there are many technical challenges to overcome. This paper will discuss several important aspects, including beamforming of cascaded channels, RIS regulation constraints, RIS system architecture for network control, integration of channel regulation and information modulation, TTD mechanisms for RIS, and so on.

*3.1. Cascade Channel Decoupling For Solving RIS Beamforming*

The introduction of RIS makes the original naturally uncontrollable electromagnetic propagation environment become a human-controllable electromagnetic propagation environment, and the active regulation of the electromagnetic propagation environment may bring a new channel paradigm. However, the cascaded channels formed after the introduction of RIS bring great challenges to the solution of the RIS beamforming matrix. The challenging analysis and a novel channel decoupling solution will be presented here. Due to space limitations, this article will only provide a brief introduction to the solution, and more detailed analysis and evaluation will be given in our other articles.

3.1.1. Challenges of Cascaded Channels

The RIS cascade channel model $H_{DL}$ given in [5] is shown as follows (ignoring the direct channel between BS-UEs). From the formula, it can be seen that the regulation coefficient matrix $\Phi$ of RIS is located between the two segmented channel matrices (the channel $G$ between NB-RIS and the channel $H$ between RIS-UE ). It is very complicated to solve the regulation coefficient matrix $\Phi$ which can best match two segmented channels simultaneously.

$$H_{DL} = H\Phi G, \qquad (1)$$

$G$, the channel between NB and RIS.

$\Phi$, the regulation coefficient matrix of RIS.

$H$, the channel between RIS and UE.

For the scenario where UE is a single antenna, literature [21] shows that the channel formula (2) can be converted into the following form.

$$H_{DL} = H\Phi G = \Phi HG, \tag{2}$$

The above expression can easily decompose $HG$ to obtain the regulation matrix $\Phi$. However, for the case where the UE is configured as multiple antennas, the above-mentioned conversion is no longer valid.

3.1.2. Channel Decoupling for Solving RIS Beamforming

However, by analyzing the regulation mechanism of RIS, it can be found that the regulation of RIS on the incident electromagnetic wave can be further divided into two sub-processes, namely, the response of the receiving sub-process to the incident electromagnetic wave and the regulation response of the outgoing sub-process. Therefore, the RIS regulation matrix $\Phi_{RIS}$ used for incident electromagnetic waves can be decomposed into two regulation matrix components corresponding to the above two sub-response processes, i.e., the receiving response matrix $\Phi_1$ and the reflection regulation matrix $\Phi_2$ at the time of the incident. The incident response of RIS to electromagnetic waves is similar to the response of the receiving beam of the analog beamforming in the Massive MIMO hybrid beamforming [22]. On this basis, the downlink received signal expression of the RIS cascaded channel is given as follows.

$$Y_{DL} = U^H H \Phi_2 \Phi_1 GFX, \tag{3}$$

$F$, the matrix of Downlink transmission beamforming at BS side,

$G$, the channel between NB and RIS,

$\Phi_1$, the receiving response matrix of RIS at the receiving sub process,

$\Phi_2$, the component of the RIS regulation matrix at the outgoing sub process,

$H$, the channel between RIS and UE,

$U^H$, the receiving beamforming matrix at the UE side.

That is to say, the RIS regulation matrix $\Phi$ is decomposed into two matrix components, $\Phi_1$ and $\Phi_2$ (i.e. $\Phi = \Phi_2 \Phi_1$), so that the solution of RIS regulation matrix $\Phi$ is transformed into the optimization of two independent components, $\Phi_1$ and $\Phi_2$.

It can be seen from the above analysis that if the singular value decomposition (SVD) mechanism is used to solve the beam forming matrix of RIS, the channel $G$ between NB-RIS can be SVD decomposed to obtain the receive matrix component $\Phi_1$ of RIS, and the channel $H$ between RIS-UE can be SVD decomposed to obtain the reflection regulation matrix component $\Phi_2$ of RIS, independently. Therefore, we can get the complete regulation matrix $\Phi$ of RIS to optimally match the G and H segmented channels at the same time. That is, **the cascaded channel is decoupled**, and the channels of $G$ and $H$ segments can be optimized independently.

Since RIS is generally deployed fixedly, the segmented channel $G$ between NB-RIS is a slowly changing channel, while the segmented channel $H$ between RIS-UE is a fast-varying channel, which is also called a time dual scale channel [23]. If the control matrix component of the RIS beamforming is decoupled and segmented, $\Phi_1$ can be calculated in a longer period and $\Phi_2$ can be calculated in a shorter one. In addition, there is no need to iteratively optimize the joint concatenated channel, which greatly reduces complexity. For example, if the codebook search mechanism is used, the codebook can be searched to match the $G$ channel with a longer period, and the codebook can be searched to match the H channel with a shorter period.

The above analysis is for the downlink (DL) MIMO channel of a single user. For the multi-user DL MIMO channel model, multiple UEs experience the same channel $G$ between BS-RIS, and the channel $H_{ris,DL}$ between RIS and multiple UEs can be expressed as follows.

$$H_{ris,DL} = diag(H_{ue,i}), \quad (4)$$

$H_{ue,i}$ the channel between RIS and $ue_i$.

Taking two UEs as examples, the downlink channel $H_{ris,DL}$ can be modified as,

$$H_{DL} = \begin{bmatrix} H_{ue1} & \\ & H_{ue2} \end{bmatrix}_{K \times M} \Phi_2 \Phi_1 G, \quad (5)$$

$K$, total number of two UE receiving antennas;
$M$, number of RIS antenna arrays.

Then, the downlink received signal is expressed as,

$$Y_{DL} = U^H \begin{bmatrix} H_{ue1} & \\ & H_{ue2} \end{bmatrix}_{K \times M} \Phi_2 \Phi_1 G F X, \quad (6)$$

Here, singular value decomposition (SVD) is used to solve $\Phi_1$ and $\Phi_2$ as follows (i.e. Block diagonalization based on singular value decomposition, BD-SVD),

$$[U, D, V_{nb-ris}^H] = SVD(G), \quad (7)$$

then, $\Phi_1 = V_{nb-ris}$.

$$[U, D, V_{ris-ue}^H] = SVD(H) = SVD \begin{bmatrix} H_{ue1} & \\ & H_{ue2} \end{bmatrix}_{K \times M}, \quad (8)$$

then, $\Phi_2 = V_{ris-ue}^H$.

*3.2. Influences and Solutions of RIS Regulatory Constraints*

From the perspective of the system model, the cascaded channel composed of NB and RIS is similar to a digital phased array plus an analog phased array of the traditional massive MIMO, realizing similar hybrid beamforming. among them, NB precoding can be compared to digital beamforming, and RIS array regulation is equivalent to analog beamforming. The spatial constraints caused by the analog beamforming of traditional massive MIMO hybrid beamforming also exist in the architecture of the cascade system composed of RIS and NB.

3.2.1. Influences of RIS Regulatory Constraints

The existing RIS without radio frequency units does not have a filtering function, and usually has broadband tuning capability (i.e., a bandwidth of several GHz) [24]. The broadband tuning characteristics of a RIS are beneficial for wireless broadband communication and support multiple bands [25]. However, each element of a RIS can only be set with a single weighting coefficient at a time, and cannot be set with different weighting coefficients for different signals on different sub-bands within the frequency range tuned by the RIS [26]. The RIS with the tuning characteristic uses the same weighting coefficient matrix to tune all signals incident on it in a wider frequency band. Therefore, the existing RIS cannot optimally match more than one sub-band channel at the same time, which may lead to serious network coexistence problems [5][6], and cannot well support multiple UEs access with FDMA.

The problem is not a new one introduced by RIS and already exists in the hybrid beamforming of massive MIMO. For analog beamforming, as the phase shifter has only one phase adjustment state, there will be only one analog beam in the system bandwidth, which makes multiple UEs with OFDMA have to share the same beamforming matrix. Different UE channels based on OFDMA are orthogonal in the frequency domain, but share the same analog beamforming matrix, which restricts the freedom of these UE channel spaces and cannot well match their channel spaces respectively. [27]

The influence of RIS tuning limitation was analyzed in our previous article [5]. Without losing generality, assuming that there is a scattering path (the channel component of $nb-ris-ue$) and a direct path (the channel component of $nb-ue$), the formula can be written as follows.

$$Y_{ue\_A} = (H_{ris\_A-ue\_A}\Theta_{ris\_A}G_{nb\_A-ris\_A} + H_{nb\_A-ue\_A})F_A X_{ue\_A} + W_{ue\_A}, \qquad (9)$$

$$Y_{ue\_B} = (H_{ris\_B-ue\_B}\Theta_{ris\_A}G_{nb\_B-ris\_A} + H_{nb\_B-ue\_B})F_B X_{ue\_B} + W_{ue\_B}, \qquad (10)$$

Where $Y_{ue\_A}$ is the received signal at $ue_A$ served by Node-B $nb_A$ of network $N_A$, $Y_{ue\_B}$ is the received signal at $ue_B$ served by Node-B $nb_B$ of network $N_B$, and $\Theta_{ris\_A}$ is the tuning coefficient matrix of $RIS_A$ for $ue_A$.

Since $RIS_A$ can only have one tuning state at a time, its coefficient matrix $\Theta_{ris\_A}$ that is suitable for the channel of $ue_A$ is also used to tune the incident signals from network $N_B$ at the same time. That is, as shown in formula (6), the tuning coefficient matrix for the signal of $ue_B$ is also $\Theta_{ris\_A}$, which leads to the deterioration of $ue_B$ performance.

*a) Impacts on Multiple Users Access*

Multi-user access adopts orthogonal frequency division multiple access. Different UE channels based on OFDMA are orthogonal in the frequency domain, but share the same RIS tuning coefficient matrix, which restricts the freedom of these UE channel spaces and cannot well match their channel spaces respectively. As shown in Figure 3, Two users UE1 and UE2 belonging to the same NB, whose downlink signals are simultaneously incident on the RIS panel, are tuned by the same matrix $\Phi_1$. The beamforming matrix $\Phi_1$ aligns with UE1, but UE2 cannot be covered by the beam, resulting in serious degradation of UE2 performance.

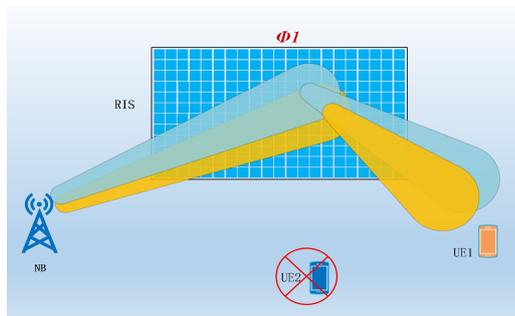

Figure 3. Impacts on Multiple Users Access.

*b) Impacts on Multiple Networks Co-existence*

As shown in the Figure 4, there are two overlapping networks, namely network A and network B (denoted by $N_A$ and $N_B$, respectively), which operate on two adjacent frequency bands. The Node-B $nb_A$ belongs to the network $N_A$, while the Node-B $nb_B$ belongs to the network $N_B$. $ue_A$ and $ue_B$ are served by $nb_A$ and $nb_B$ respectively. The RIS $RIS_A$ of the network $N_A$ tunes the signal from the network $N_A$ using the coefficient matrix $\Phi_1$ based on the channel of $ue_A$. $RIS_A$ will also tune the signal from the network $N_B$ at the same time using the same coefficient matrix $\Phi_1$ for the signal of network $N_A$. So, the tuning of $RIS_A$ causes a unexpected disturbance on the channel of the non-target signal from network $N_B$.

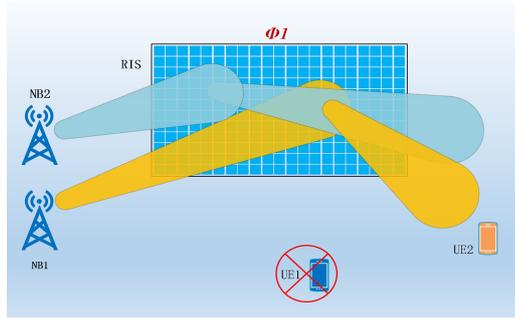

Figure 4. Impacts on Multiple Networks co-existence.

3.2.2. Solutions for RIS Regulatory Constraints

In this section, two novel RIS design mechanisms proposed by us [5][6], including a novel multi-layer RIS structure with an out-of-band filter and a RIS blocking mechanism, are further explored. Among them, the former adopts the idea of eliminating influence, while the latter adopts the idea of reducing influence. The advantages and possible negative effects of these two mechanisms are comprehensively analyzed and evaluated.

*a) **RIS BLOCKING SCHEMES***

In [5][6], we proposed a RIS blocking mechanism to solve problems of the RIS network coexistence. The basic idea of the RIS blocking mechanism is that the incident signals of different UEs can be assigned to different sub-blocks of the RIS, and each sub-block can tune signals incident on it by using an independent coefficient matrix separately, as shown in Figure 5. It is also possible to block the antenna array elements of RIS in the way of interval decimation for grouping (as shown in Figure 6). Here, the preferred way is that the RIS adopts more dense antenna elements, so that the interval of antenna elements of each sub-group after grouping still meets the conditions of less than or equal to $\lambda/2$.

From the perspective of the UE source, the mechanism is used for multi-UE scheduling when the UEs come from the same network. The mechanism is used for RIS network coexistence when the UEs come from different networks. For the RIS network coexistence scenario, the RIS can only be divided in a static or semi-static way, since it is difficult to dynamically coordinate between Node-Bs, especially Node-Bs coming from different operators. It is necessary to deploy an appropriate RIS antenna size and optimize a reasonable blocking ratio for different networks to ensure the network's own performance as much as possible while satisfying the coexistence performance.

Without loss of generality, assuming that a RIS is divided into two sub-blocks and that the $ue_A$ signal is set as the target signal, the formula can be written as follows.

$$Y_{ue\_A} = \sqrt{\beta}(H_{ris\_sub1-ue\_A}\Theta_{ris\_A}G_{nb\_A-ris\_sub1} + H_{nb\_A-ue\_A})F_A X_{ue\_A} + \sqrt{(1-\beta)}(H_{ris\_sub2-ue\_A}\Theta_{ris\_B}G_{nb\_A-ris\_sub2})F_A X_{ue\_A} + W_{ue\_A} \quad (11)$$

Where $Y_{ue\_A}$ is the received signal of $ue_A$ served by base station $nb_A$, $\beta$ is the energy proportion of the $ue_A$ signal incident on its RIS sub-block, $(1-\beta)$ is the energy proportion of the $ue_A$ signal incident on another RIS sub-block, $\Theta_{ris\_A}$ is the optimal tuning matrix of the RIS sub-block for the $ue_A$ signal, and $\Theta_{ris\_B}$ is the optimal tuning matrix of the RIS sub-block for the $ue_B$ signal.

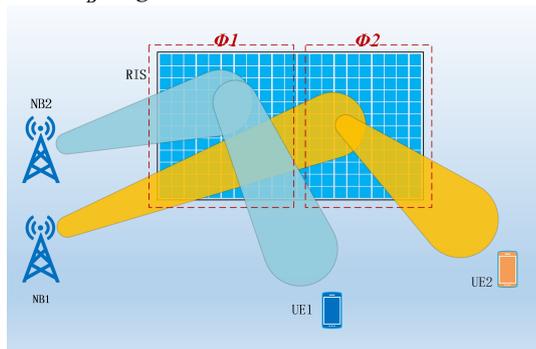

**Figure 5**. Tuning signals incident on a sub-block by using an independent coefficient matrix separately.

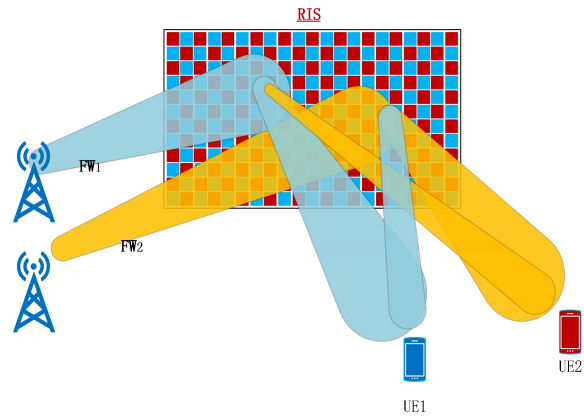

**Figure 6**. Block the antenna array elements of RIS in the way of interval decimation for grouping.

*b) Multiple-layers RIS with filter layer(s)*

In our previous articles [5][6], a novel RIS structure, multiple-layer RIS with filter layers, was also proposed. Without losing generality, let us take a RIS with a double-layer meta-surface structure as an example. The first layer of the RIS is a bandpass filter using a meta-surface. The bandpass filter only allows signals in the target band to pass through, while the signals in the adjacent non-target bands (out-of-band signals) are filtered. The second layer of the RIS is a conventional programmable meta-surface that can realize typical programmable functions of RIS. The programmable meta-surface will only tune the target signal, since the non-target signal has been filtered by the first layer, as shown in Figure 7.

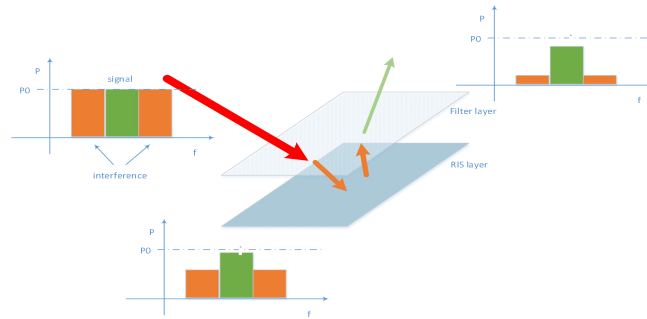

**Figure 7**. Out-of-band filtering using an RIS with a double-layer structure (reflecting the RIS) [5]

The expression of the $ue_A$ signal served by $NB_A$ and the $ue_B$ signal served by $NB_B$ can be modified from formula (3), as shown below.

$$Y_{ue\_A} = (H_{ris\_A-ue\_A}\Theta_{ris\_A}G_{nb\_A-ris\_A} + H_{nb\_A-ue\_A})F_A X_{ue\_A} + (H_{ris\_B-ue\_A}\beta\Theta_{ris\_B}G_{nb\_A-ris\_B})F_A X_{ue\_A} + W_{ue\_A} \quad (12)$$

$$Y_{ue\_B} = (H_{ris\_B-ue\_B}\Theta_{ris\_B}G_{nb\_B-ris\_B} + H_{nb\_B-ue\_B})F_B X_{ue\_B} + (H_{ris\_A-ue\_B}\beta\Theta_{ris\_A}G_{nb\_B-ris\_A})F_B X_{ue\_B} + W_{ue\_B} \quad (13)$$

Where, $Y_{ue\_A}$ refers to the received signal of $ue_A$ served by the base station $nb_A$, $\beta \in (0,1)$ refers to the filter coefficient of $ris_A$, $\Theta_{ris\_A}$ refers to the optimal tuning matrix of $ris_A$ for the signal of $ue_A$, and $\Theta_{ris\_B}$ refers to the optimal tuning matrix of $ris_B$ for the signal of $ue_B$.

*3.3. RIS System Architecture of Network Controlled Mode*

Most existing RIS studies assume that there is a controller connected to the network to control the electromagnetic regulation behavior of RIS. Especially for the multi-RIS scenarios, a controller is needed to control the cooperation among multiple RISs. For a RIS deployed in the cellular network, the RIS regulates electromagnetic waves under the control of the network. The RIS control mode can achieve better network performance, while it also brings the complexity to controlling information interaction.

Recently, to overcome the complexity of control information interaction between RIS and the network, literature [28][29] proposed that RIS self-control mode can be adopted. Literature [29] proposed that RIS can adopt a blind beamforming mechanism to achieve self-regulation and beamforming. However, the mechanism of blind beamforming proposed in the literature is based on statistical channel information, which is inferior to beamforming of knowing CSI. The blind beamforming is only suitable for single-user scenarios. Literature [28] proposed that RIS can self-implement channel estimation and beamforming based on interference phenomena. Again, this mechanism can only be used in single-user scenarios.

Whether it is interference-based channel estimation and beamforming or blind beamforming mechanism, it is necessary to assume that RIS has to know which UE to be served, the target to be optimized, and be in a single UE scenario. In addition, it is necessary for the NB or other network units to know and inform the RIS of the above information, or to complicate the RIS so that the RIS has the functions like the NB. These two mechanisms mentioned above are difficult to realize coordination and optimization among multiple users and multiple networks.

From the above analysis, the above two mechanisms enable RIS to have certain distributed computing capabilities, which can reduce NB's demand for centralized computing, and thus reduce the interaction between the NB and RIS. However, they cannot completely cancel the interaction between RIS and the network. Table 4 compares the advantages and challenges of independent deployment mode and network control deployment mode. Table 5 shows the requirements and candidate solutions when adopting the network control deployment mode.

**Table 4**: Advantages and challenges of network control mode and Standalone mode.

| Type | Advantages | Challenges |
|---|---|---|
| **Network-Controlled Mode** | ① Support multi-network collaboration. ② Supports multi-user access. ③ Better meet the coexistence requirements of wireless networks deployed on licensed spectrum. | ① Network deployment is relatively complicated. ② Network control link needs to be deployed. ③ It is necessary to design an interactive flow of control and measurement signaling. |
| **Standalone Mode** | ① No network control link is required; ② The network is simple and easy to deploy; ③ suitable for unlicensed spectrum with low coexistence requirements. | ① Hard to overcome the interference of multiple networks. ② May cause serious inter-cell interference. ③ Cannot support multi-user access well. |

**Table 5**: Requirements and candidate solutions of network control mode.

| | Requirements | Candidate solutions |
|---|---|---|
| **Feedback of CSI and Scheduling of Beamforming** | ① The number of super large antenna arrays brings huge pilot cost and channel information feedback cost | ① Utilize channel sparsity, adopt compression sensing and other mechanisms to reduce overhead ② Using statistical channel |

|  | ② Real-time demand of channel measurement and feedback. | information<br>③ Using the Dual Time Scale Property of Cascaded Channels<br>④ Beam scanning and codebook quantization. |
|---|---|---|
| **Network control link** | The control link needs to be specially deployed, resulting in the complexity and cost of deployment. Especially in remote areas, complexity and cost are very big problems. | Flexible choice of wired or wireless links. |
| **Interactive flow of control and measurement signaling** | The protocol standardization of Interactive flow of control and measurement signaling is necessary. | The related protocol flow can be standardized in the future 5G-A or 6G standards. [30] |

Based on the comparison and analysis of the above two tables, combined with the characteristics of licensed spectrum and unlicensed spectrum, the following conclusions can be drawn.

(1) Network-controlled mode: suitable for scenarios such as complex networks and licensed spectrum with high coexistence requirements (i.e., cellular networks).

(2) Standalone mode: suitable for scenarios such as simple networks, unlicensed spectrum technology with local coverage, and so on (e.g., Wi-Fi).

*3.4. Integrated Channel Regulation and Information Modulation*

Different from the current popular research (RIS is only used to assist the communication of existing communication systems), a RIS function called symbiotic radio (SR) transmission has recently been proposed [31]. RIS used as SR not only help the transmitter to enhance the transmission of the primary wireless network through passive beamforming, but also transmits its own information to receivers by using reflected signals. This concept is similar to spatial modulation transmission, in which the indices of the active transmitting antenna are used to encode information, to improve spectral efficiency. The RIS acts not only as an information source node but also as a helper for improving the performance of the primary link via passive beamforming. Comparing several similar technologies, we can find that in essence, whether backscatter [10], spatial modulation [32], or information meta-surface [33], information modulation is realized by amplitude/phase modulation of the carrier.

3.4.1. Requirements and challenges

The existing researches focus on using RIS to provide information modulation and transmission functions at the same time, which is used to realize the functions of traditional backscatter technology to support passive IoT communications. But up to now, there is no literature to study the RIS-based information modulation and transmission to support the electromagnetic wave regulation function of RIS itself, that is, to support the control and measurement information interaction between RIS-NB and RIS-UE.

The ancillary functions supported by RIS-based backscatter include (1) identifying RIS IDs for multiple RIS deployment scenarios. It can be used for NB/UE to discover the existence of surrounding RIS, and obtain the location and characteristic parameters of RIS, to effectively use RIS; (2) used for transmitting measurement and control information, such as RIS control signaling/CSI; (3) RIS is also used for electromagnetic environment regulation and information modulation similar to backscatter.

Since one RIS symbol is transmitted in K successive legacy symbol period, the RIS-based information modulation can be regarded as a spread-spectrum code of the RIS symbols for large K. E.g., repeating in time/frequency domain; Spreading in the spatial domain (antenna domain, i.e. different antennas).

Since the data rate of measurement and control information exchanged between RIS and NB is relatively low, the data rate that RIS-based backcatcher can achieve should meet the above information exchange requirements. However, there are still some challenges to better support the exchange of measurement and control information and ensure the performance of the primary system (traffic transmission between NB and UE). Here, the main challenges that may be faced are analyzed and preliminary solutions are given.

**(1) Challenge 1**

RIS is used as a backscatter to modulate information, resulting in random fluctuations in the regulation coefficient originally used for channel regulation by RIS, that is, RIS introduces additional noise. This effect is similar to the traditional active-phased antenna or relay, in which the analog transmitter introduces phase (amplitude) noise. Because the modulation of the information bit is random, the noise is random. The unexpected fluctuation of the channel caused by the RIS-based information modulation results in inaccurate CSI estimation, and the modulation also leads to the disturbance of the phase and amplitude of the beamforming coefficient, which was originally used for channel regulation of the primary system. This will cause two problems: a) CSI measurement estimation error; b) Beam regulation error during data transmission of the primary system. For example, RIS uses the ON/OFF modulation mode to modulate information. In the OFF slot, the primary system cannot transmit information, resulting in a reduction in the available time domain resources of the primary system. When the ON-OFF regulator makes information modulation, the capacity loss of the main transmitter is reflected in the power loss due to the OFF state. During the OFF state, the signal energy of the main transmitter is 0, which leads to the reduction of its effective power, SNR and capacity loss. Alternatively, it can be reduced from the proportion of the channel available time, resulting in the reduction of the theoretical capacity. For infinite-length coding, the OFF ratio can be converted into the reduction ratio of the total channel capacity. In other words, the OFF state is unavailable, resulting in the loss of time domain resources, which leads to the reduction of channel capacity.

**(2) Challenge 2**

This part discusses the interference of the direct channel. The signal of the direct channel will interfere with the RIS modulation information, especially in the scenario where the direct channel signal is strong. The receiver of the backscatter communication (BC) always encounters the strong direct-link interference of the RF source. To solve this problem, various interference cancellation techniques have been developed [34][35][36].

**(3) Challenge 3**

RIS-based information modulation and transmission depend on the existence of the primary system signal, and the transmission direction of the primary system signal needs to be considered. When the traffic load of the primary system is light, or there is almost no signal transmission on the downlink/uplink, it is difficult for BC to use the carrier resources of the primary system to modulate and transmit information in time, resulting in low information rate and large delay. To solve this problem, the possible solution is to add additional reference signals, which can be used to supplement and enhance the measurement requirements of the primary system on the one hand, and assist the backscatter device (BD) information modulation and transmission requirements on the other hand, such as sending additional dedicated downlink reference signals, UE uplink Sounding RS and so on.

**(4) Challenge 4**

Traditional RIS beamforming optimization is generally aimed at the NB/UE of the primary system. If the BC receiver and NB/UE are not integrated, the RIS beam will probably not be aligned with the BC receiver, resulting in poor signal quality of the BC receiver. For BC to receive its signal better, RIS needs to optimize beamforming and align the BC receiver, which will lead to the performance degradation of the primary system, especially in the high-frequency narrow-beam scenarios.

**(5) Challenge 5**

When RIS is used as a backscatter, its behavior may have a significant impact on the communication of the primary system, since its antenna aperture is large and usually deployed on the main propagation path of the primary system. If it obtains its regulated energy source through wireless energy collection, it will have a greater influence on the signal energy of the primary system path, resulting in a greater impact on the performance of the primary system.

Some solutions can be considered for the challenge. (a) RIS is divided into blocks. Only some sub-blocks are used for the backcatcher functions by reducing the original unique functions of RIS for the primary system. (b) Some joint optimization mechanisms can be considered to optimize both the functions of RIS to the primary system and the information modulation function/energy collection function of backcatcher. Note, typical backscattering devices are generally small in size, so the impact on the primary system may not be significant.

3.4.2. Solution: A Novel Frame Structure

A novel frame structure is provided here to realize that the RIS supports both channel regulation and information modulation, as shown in Figure 8.
   a) Pilot time slot, not modulate the information and only regulate the channel of the primary system.
   b) Backscatter slot, do the channel regulation and information modulation simultaneously.
   c) It is better to have RS signals in both time slots, to estimate the channels of the two types of time slots respectively. The difference between these two channels is the modulated information. By using RS, we can estimate these two channels more accurately.
   d) The design of the frame structure needs to consider time domain correlation, overhead, RS structure, etc.
   e) In addition, RIS can consider designing a silent time slot/frame, so that only the signals of the direct channels/signals can be detected in these time slots/frames and the detection performance of these signals can be guaranteed.

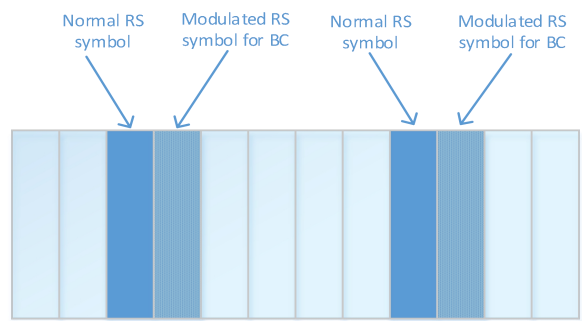

Figure 8. A Novel Frame Structure for a RIS Integrated Channel Regulation and Information Modulation.

*3.5. True Time Delay (TTD) Schemes for RIS*

When studying the analog beamforming in the massive MIMO hybrid beamforming, it is found that the beam splitting[2] ("squint" phenomenon) will occur in different frequency sub-bands under the broadband, which is also called the rainbow effect. To solve this phenomenon, the researcher proposed to use the True Time Delay (TTD) mechanism to eliminate the phase deflection difference of different sub-bands caused by time delay [37][38][39]. Literature [40] further pointed out that TTD can also be used to solve the problem of the "squint" phenomenon faced by RIS.

TTD mechanism uses the precise adjustment and control of delay to align the offset phase difference of different frequency sub-bands. Naturally, by adjusting different delay differences, this technical idea can be used to generate independent phase adjustment effects for different frequency sub-bands. Literature [41] hopes to give the frequency-dependent analog beamforming by using TTD. However, this research is aimed at the analog beamforming problem of large-scale MIMO, which is different from the requirements of RIS. That is, RIS and Massive MIMO active phased arrays have different adjustment mechanisms and face different implementation limitations. It is necessary to deeply study the mechanism of RIS scenarios to realize independent Analog Beamforming in different sub-bands based on TTD, which can be used to support the requirement that different UEs with OFDMA independently adopt their own beamforming in their own occupied sub-bands.

3.5.1. Concepts of TTD Scheme

In principle, the TTD circuit introduces a group delay to the received signal. The delay is Denoted as $\tau_{TTD,n}$ in the $n-th$ array element, then the received OFDM symbol of the $m-th$ sub-carrier is

$$Y_m = w_{TTD,m}^H H_m v X_m + N_m, \qquad (14)$$

The combiner specified by TTD arrays $w_{TDD,m} \in C^{N_R}$ is frequency dependent, i.e., its $n-th$ element as

$$w_{TTD,m,n}^H = e^{j2\pi f_m \tau_{TTD,n}}, \tag{15}$$

where, the RF frequency of the $m-th$ subcarrier is denoted as $f_m$, the channel at the $m-th$ subcarrier is denoted as $H_m$.

---

[2] The concept of the beam angle changing as a function of frequency is called beam squint.

This model is valid as long as CP is longer than the propagation delay $\Gamma_{l,q,n}$ and the cumulative delay $\tau_{TTD,n}$ of TTD circuit, i.e.,

$$N_{cp}T_s > \max_{l,q,n} \Gamma_{l,q,n} + \max_{l,q,n} \tau_{TTD,n}, \tag{16}$$

The frequency domain noise $N_m$ is Gaussian distributed.

3.5.2. RIS Using TDD to Support Frequency-Dependent Beamforming for Multiple Users Access with OFDMA

As mentioned above, it is assumed that RIS can only have one phase/amplitude modulation weighted mode at the same time, so the general beamforming can only best match the channel of a UE. In the scenario of multi-user OFDMA access, that is, different UEs use different sub-bands in the FDM mode, RIS will not be able to best match the frequency orthogonal channels of these UEs respectively.

In this section, we will discuss the application of the TTD mechanism in RIS to support the frequency-independent beamforming for different UEs with OFDMA in their respective occupied sub-bands. TTD mechanism is flexible in phase adjustment of different sub-bands, that is, it increases phase freedom in the frequency domain, which can bring scheduling performance gains.

Take the case where two UEs ($ue_1$ and $ue_2$) use OFDMA as an example. $ue_1$ occupies the sub-band $B_1$, and $ue_1$ occupies the sub-band $B_2$. The precoding matrix $W_1$ of $ue_1$ is calculated based on the channel $H_1$ of $ue_1$ corresponding to the center frequency point $f_1$ of $B_1$. The precoding matrix $W_2$ of $ue_2$ is calculated based on the channel $H_2$ corresponding to the central frequency point $f_2$ of $B_2$. Then, the phase difference vector $\Phi$ corresponding to each antenna port corresponding to $W_1$ and $W_2$, where $\phi_n$ represents the phase difference between $W_1$ and $W_2$ corresponding to the antenna $n-th$.

The delay difference corresponding to the phase difference can be calculated by the following formula. That is, the time delay difference is inversely calculated according to the phase difference.

$$w_{m,n} = \alpha_n \exp(-j(2\pi(f_m - f_c)\tau_n + \phi_n)), \tag{17}$$

Then the TDD coefficient matrix of m-subcarrier is $W_m = diag(w_{m,n}), n = 1,...,N$.

Basic process of implementing TDD mechanism.

(1) taking sub-band 1 (center frequency $f_1$) as the reference frequency $f_c = f_1$, the RIS regulation matrix $\Phi_{f1} = \{\phi_{f1,1},...,\phi_{f1,N}\}$ corresponding to $f_1$ is calculated by the traditional method, $N$ is the number of antennas. Then, the phase corresponding to the $m$-$th$ sub-carrier of the antenna $n$-$th$ in sub-band2 (central frequency point $f_2$).

(2) after adding time delay $\tau_n$, additional phase offset $\Delta\phi_n$ will be introduced at $f_1$, so the original regulation matrix $\Phi_{f1}$ needs to be revised, i.e., $w'_{f1\_n} = w_{f1} * e^{-j\Delta\phi_n}$.

(3) Because the phase difference between the two sub-bands is realized based on the time delay, the beam squint will be introduced into all sub-carriers except the center frequency points of BW 1 and BW 2 due to the time delay $\tau_n$. Especially when the delay $\tau_n$ is large (because of the large phase difference $\phi_n$, it is necessary to set a large delay $\tau_n$ accordingly), the beam squint effect will be significant, resulting in performance degradation. To minimize the impact of the beam squit mentioned above, the following factors should be considered when optimizing the algorithm, (a) optimization objectives: maximize the sum capacity of multiple subbands; (b) at least several constraints should be considered, i.e., the range of allowable delay $\tau_n$, the performance impact caused by beam squint (e.g., compared with the TTD used to perfectly eliminate beam squint), and the minimum traffic delay (such as URLLC); (c) During multi-user scheduling, it should be considered that the beam phase difference of different UEs should not be too large to minimize the influence of beam squint.

(4) Other considerations, including number of delay/phase quantization bits, e.g., 1~2 bits; considering the complexity and cost; the range of time delay/phase, and so on.

(5) In order to reduce complexity, $K$ adjacent antenna elements of RIS can be grouped and share a time delay.

## 4. Future Trends and Challenges

To further promote the industrialization process of RIS, some aspects need further in-depth research and industrial promotion.

*4.1. A New Structure RIS: RIS Integrated with Photovoltaic Panels*

As mentioned above, RIS is expected to be widely deployed in future wireless networks, which will bring about power supply problems, especially in remote areas. However, RIS is a plate structure with low power consumption. It is a natural idea to consider that RIS integration can also supply power for photovoltaic panels with plate structures, especially for outdoor scenarios.

There are two possible ways for RIS Integrating Photovoltaic Panels.

Mode 1, a novel RIS structure is designed, that is, the optically transparent RIS is located on the surface of the photovoltaic panel to regulate wireless electromagnetic waves; The energy of visible light falls on the surface of the photovoltaic panel through RIS, and generates electricity to support the RIS control unit, and even meet the power demand of other surrounding devices.

Mode 2, Additional photovoltaic panels are deployed for supplying power to RIS. Because both of them are plate-like structures, it is easy to design composite structures. This model is not a new RIS structure, but a new combination deployment model.

Of course, the integration structure also faces some challenges. For example, photovoltaic power generation can only have high power when the sunlight is strong in the daytime. However, since data traffics is mainly in the daytime and at night, there is almost no service in the middle of the night, so the power consumption is low. Therefore, the peak of business during the day just corresponds to the peak of photovoltaic power generation, and the energy storage module is needed to support it at night. Photovoltaic panels and RIS panels have different requirements for deployment angles, which brings difficulties in integrated design. Other challenges include low light on cloudy days, the possibility of relatively low transmittance of RIS, and so on.

*4.2. Complexity of RIS Network Deployment and Optimization*

The introduction of RIS will bring a new network paradigm to the future network, but it will also lead to the complexity of network deployment and optimization, including the following.

① Multiple transmission nodes share the same RIS.

② Collaborative scheduling when a transmission node uses multiple RIS at the same time.

③ control link between transmission node and RIS (such as. NB-RIS return link).

④ The challenge faced by network topology planning and optimization. RIS enhances and expands the signal coverage area, which may break the traditional strictly divided sector coverage characteristics and bring complexity to network planning and optimization.

⑤ Station site selection and power supply. The simplicity, low cost and low power consumption of RIS make it possible to deploy RIS more widely. However, its ubiquitous deployment may also bring new challenges to RIS power supply and management.

⑥ different scenarios need different forms of RIS, so it is necessary to design different forms of RIS.

*4.3. Engineering Errors*

Due to the limitation of cost and complexity, the engineering implementation of RIS will have some engineering errors. The influence of these engineering errors should be considered in the algorithm design and optimization. The main engineering deviations of RIS can include: ① control phase quantization error, for example, 1-bit quantization, 2-bit quantization; ② deviation calibration, including processing deviation and calibration, aging, environmental impact, and other system drift measurement calibration; ④ CSI feedback quantization error and time delay, etc.

*4.4. Standard and Protocol Design*

The standardized design of RIS mainly involves the following aspects.

1) the access procedure of UE. The access procedure design includes the measurement, discovery, random access, transmission, target signal identification and regulation, measurement, and so on.

2) resource management and scheduling procedure. It is mainly for the management and scheduling of RIS resources, such as the selection of RIS channels and NB-UE direct channels, and the scheduling/selection of RIS in multiple RIS scenarios.

3) the impact of the introduction of RIS on the handover procedure (i.e. mobility management). Such as switching between different RIS, switching between different frequency bands, etc.

4) Side control information interaction between NB and RIS, such as beamforming information, timing information to align transmission, information on UL-DL TDD configuration, ON-OFF information, etc.

## 5. Conclusion

This paper focuses on aspects of the actual engineering applications of RIS The engineering applications of RIS were summarized. And then, technical challenges and solutions of RIS engineering applications were discussed. Future trends and challenges are also provided. We can conclude that, although the engineering applications of RIS face many challenges, there are corresponding solutions to overcome these challenges.

**References**

1. Zhao Y J, Yu G H, Xu H Q. 6G mobile communication networks: vision, challenges, and key technologies (in Chinese). *Sci Sin Inform*, **2019**, *49*: 963-987, doi: 10.1360/N112019-00033.
2. Yuan Y F, Zhao Y J, Zong B Q, et al. Potential key technologies for 6G mobile communications. *Sci China Inf Sci*, **2020**, *63(8)*: 183301, https://doi.org/10.1007/s11432-019-2789-y.
3. NTT DOCOMO and Metawave Announce Successful Demonstration of 28GHz-Band 5G Using World's First Meta-Structure Technology. Available online: https://www.businesswire.com/news/home/20181204005253/en (December 04, 2018)
4. Ma H B, Zhang P, Yang F, et al. Reflections on Reconfigurable Intelligent Surface Technology (in Chinese). *ZTE Technology Journal*, **2022**, *28(03)*:70-77.
5. Y. Zhao and X. Lv, "Network Coexistence Analysis of RIS-Assisted Wireless Communications," in IEEE Access, vol. 10, pp. 63442-63454, 2022, doi: 10.1109/ACCESS.2022.3183139.
6. Y. Zhao and M. Jian, "Applications and challenges of reconfigurable intelligent surface for 6G networks", *arXiv*:2108.13164, **2021**.
7. L. Lu, G. Y. Li, A. L. Swindlehurst, A. Ashikhmin and R. Zhang, "An Overview of Massive MIMO: Benefits and Challenges," in IEEE Journal of Selected Topics in Signal Processing, vol. 8, no. 5, pp. 742-758, Oct. 2014, doi: 10.1109/JSTSP.2014.2317671.
8. H. Xu, Y. Zhao, L. Mo, C. Huang and B. Sun, "Inter-cell antenna calibration for coherent joint transmission in TDD system," 2012 IEEE Globecom Workshops, 2012, pp. 297-301, doi: 10.1109/GLOCOMW.2012.6477586.
9. D. Darsena, G. Gelli and F. Verde, "Design and performance analysis of multiple-relay cooperative MIMO networks," in Journal of Communications and Networks, vol. 21, no. 1, pp. 25-32, Feb. 2019, doi: 10.1109/JCN.2019.000003.
10. N. Van Huynh, D. T. Hoang, X. Lu, D. Niyato, P. Wang and D. I. Kim, "Ambient Backscatter Communications: A Contemporary Survey," in IEEE Communications Surveys & Tutorials, vol. 20, no. 4, pp. 2889-2922, Fourthquarter 2018, doi: 10.1109/COMST.2018.2841964.
11. ZTE, RP-213700 New SI: Study on NR Network-controlled Repeaters, , 3GPP TSG RAN Meeting #94e, Dec. Electronic Meeting, 6 - 17, 2021.
12. Z. Zhang, L. Dai, X. Chen, C. Liu, F. Yang, R. Schober, and H. V. Poor, "Active RIS vs. passive RIS: Which will prevail in 6G?," arXiv Preprint, arXiv:2103.15154, Mar. 2021.
13. Y. Liu, X. Mu, R. Schober and H. V. Poor, "Simultaneously Transmitting and Reflecting (STAR)-RISs: A Coupled Phase-Shift Model," ICC 2022 - IEEE International Conference on Communications, 2022, pp. 2840-2845, doi: 10.1109/ICC45855.2022.9838767.
14. S. Guo, S. Lv, H. Zhang, J. Ye and P. Zhang, "Reflecting Modulation," in IEEE Journal on Selected Areas in Communications, vol. 38, no. 11, pp. 2548-2561, Nov. 2020, doi: 10.1109/JSAC.2020.3007060.
15. T. J. Cui, "Information Metamaterial and Metasurface - From Concept to System," 2018 43rd International Conference on Infrared, Millimeter, and Terahertz Waves (IRMMW-THz), 2018, pp. 1-3, doi: 10.1109/IRMMW-THz.2018.8510031.
16. Z. Zhang, L. Dai, X. Chen, C. Liu, F. Yang, R. Schober, and H. V. Poor, "Active RIS vs. passive RIS: Which will prevail in 6G?," arXiv Preprint, arXiv:2103.15154, Mar. 2021.
17. A. Taha, M. Alrabeiah and A. Alkhateeb, "Enabling Large Intelligent Surfaces With Compressive Sensing and Deep Learning," in IEEE Access, vol. 9, pp. 44304-44321, 2021, doi: 10.1109/ACCESS.2021.3064073.

18. Y. Zhang, K. Shen, S. Ren, X. Li, X. Chen and Z. -Q. Luo, "Configuring Intelligent Reflecting Surface With Performance Guarantees: Optimal Beamforming," in IEEE Journal of Selected Topics in Signal Processing, vol. 16, no. 5, pp. 967-979, Aug. 2022, doi: 10.1109/JSTSP.2022.3176479.
19. J. Zhu, K. Liu, Z. Wan, L. Dai, T. J. Cui, and H. V. Poor, "Sensing RISs: Enabling dimension-independent CSI acquisition for beamforming,"arXiv preprint arXiv:2204.13505., Apr. 2022.
20. Zhao Y J, Zhang J Y, Ai B. Applications of Reconfigurable Intelligent Surface in Smart High Speed Train Communications (in Chinese). *ZTE Technology Journal,* **2021***, 27(04):36-43.*
21. X. Wei, D. Shen and L. Dai, "Channel Estimation for RIS Assisted Wireless Communications—Part II: An Improved Solution Based on Double-Structured Sparsity," in IEEE Communications Letters, vol. 25, no. 5, pp. 1403-1407, May 2021, doi: 10.1109/LCOMM.2021.3052787.
22. G. Zhu, K. Huang, V. K. N. Lau, B. Xia, X. Li and S. Zhang, "Hybrid Beamforming via the Kronecker Decomposition for the Millimeter-Wave Massive MIMO Systems," in IEEE Journal on Selected Areas in Communications, vol. 35, no. 9, pp. 2097-2114, Sept. 2017, doi: 10.1109/JSAC.2017.2720099.
23. C. Hu, L. Dai, S. Han, and X. Wang, "Two-timescale channel estimation for reconfigurable intelligent surface aided wireless communications," IEEE Trans. Commun., vol. 69, no. 11, pp. 7736-7747, Nov. 2021.
24. D. Wang, Y. Tan, L.-Z. Yin, T.-J. Huang and P.-K. Liu, "A subwavelength 1-bit broadband reconfigurable reflectarray element based on slotting technology", Proc. Int. Appl. Comput. Electromagn. Soc. Symp.-China (ACES), pp. 1-2, Aug. 2019.
25. T.-J. Cui, H.-T. Wu and S. Liu, "Research progress of information metamaterials",?Acta Phys. Sinica, vol. 69, no. 15, 2020.
26. N. Avazov, R. Hicheri, M. Muaaz, F. Sanfilippo and M. Patzold, "A trajectory-driven 3D non-stationary mm-wave MIMO channel model for a single moving point scatterer",?IEEE Access, vol. 9, pp. 115990-116001, 2021.
27. R. Rotman, M. Tur and L. Yaron, "True Time Delay in Phased Arrays," in Proceedings of the IEEE, vol. 104, no. 3, pp. 504-518, March 2016, doi: 10.1109/JPROC.2016.2515122.
28. J. Zhu, K. Liu, Z. Wan, L. Dai, T. J. Cui, and H. V. Poor, "Sensing RISs: Enabling dimension-independent CSI acquisition for beamforming,"arXiv preprint arXiv:2204.13505., Apr. 2022.
29. Y. Zhang, K. Shen, S. Ren, X. Li, X. Chen and Z. -Q. Luo, "Configuring Intelligent Reflecting Surface With Performance Guarantees: Optimal Beamforming," in IEEE Journal of Selected Topics in Signal Processing, vol. 16, no. 5, pp. 967-979, Aug. 2022, doi: 10.1109/JSTSP.2022.3176479.
30. M. Jian et al., "Reconfigurable intelligent surfaces for wireless communications: Overview of hardware designs, channel models, and estimation techniques," in Intelligent and Converged Networks, vol. 3, no. 1, pp. 1-32, March 2022.
31. M. Hua, Q. Wu, L. Yang, R. Schober and H. V. Poor, "A Novel Wireless Communication Paradigm for Intelligent Reflecting Surface Based Symbiotic Radio Systems," in IEEE Transactions on Signal Processing, vol. 70, pp. 550-565, 2022, doi: 10.1109/TSP.2021.3135603.
32. E. Basar, "Reconfigurable Intelligent Surface-Based Index Modulation: A New Beyond MIMO Paradigm for 6G," in IEEE Transactions on Communications, vol. 68, no. 5, pp. 3187-3196, May 2020, doi: 10.1109/TCOMM.2020.2971486.
33. T. J. Cui, "Information Metamaterial and Metasurface - From Concept to System," 2018 43rd International Conference on Infrared, Millimeter, and Terahertz Waves (IRMMW-THz), 2018, pp. 1-3, doi: 10.1109/IRMMW-THz.2018.8510031.
34. D.Bharadia, K. R. Joshi, M.Kotaru, and S.Katti, " Backfi: High throughput WiFi backscatter, " ACMSIGCOMMComput.Commun. Rev., vol. 45, no. 4, pp. 283–296, Aug. 2015.
35. B. Kellogg, V. Talla, S. Gollakota, and J. R. Smith, "Passive Wi-Fi: Bringing low power to Wi-Fi transmissions," in Proc. 13th USENIX Symp. Networked Syst. Des. Implementation. Santa Clara, CA: USENIX Association, 2016, pp. 151–164.
36. G. Yang, Y.-C. Liang, R. Zhang, and Y. Pei, "Modulation in the air: Backscatter communication over ambient OFDM carrier," IEEE Trans. Commun., vol. 66, no. 3, pp. 1219–1233, Mar. 2018.
37. H. Yan, V. Boljanovic and D. Cabric, "Wideband Millimeter-Wave Beam Training with True-Time-Delay Array Architecture," 2019 53rd Asilomar Conference on Signals, Systems, and Computers, 2019, pp. 1447-1452, doi: 10.1109/IEEECONF44664.2019.9048885.
38. S. -H. Park, B. Kim, D. Ku Kim, L. Dai, K. -K. Wong and C. -B. Chae, "Beam Squint in Ultra-wideband mmWave Systems: RF Lens Array vs. Phase-Shifter-Based Array," in IEEE Wireless Communications, doi: 10.1109/MWC.007.2100530.
39. Chen, Yihong, K. Wu, Feng Zhao, Gicherl Kim and Ray T. Chen. "Reconfigurable true-time delay for wideband phased-array antennas." SPIE OPTO (2004).
40. Lin, C., Boljanovic, V., Yan, H., Ghaderi, E., Mokri, M.A., Gaddis, J.J., Wadaskar, A., Puglisi, C., Mohapatra, S., Xu, Q., Poolakkal, S., Heo, D., Gupta, S., & Cabric, D. . Wideband Beamforming with Rainbow Beam Training using Reconfigurable True-Time-Delay Arrays for Millimeter-Wave Wireless. *ArXiv,* abs/2111.15191. **2021**.
41. V. V. Ratnam, J. Mo, A. Alammouri, B. L. Ng, J. Zhang and A. F. Molisch, "Joint Phase-Time Arrays: A Paradigm for Frequency-Dependent Analog Beamforming in 6G," in *IEEE Access*, *vol. 10*, pp. 73364-73377, **2022**, doi: 10.1109/ACCESS.2022.3190418.

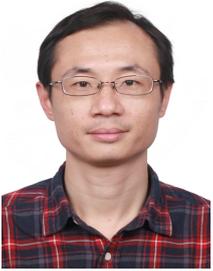

**Yajun Zhao** has B.E. and master's degrees. Since 2010, he has acted as a radio expert in the wireless advanced research department at the ZTE Corporation. Prior to this, he worked for Huawei on wireless technology research in the wireless research department. Currently, he is mainly engaged in research on 5G standardization technology and future mobile communication technology (6G). His research interests include reconfigurable intelligent surface (RIS), spectrum sharing, flexible duplex, CoMP, and interference mitigation.

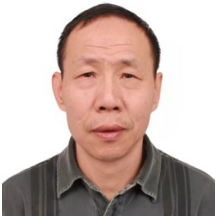

**Xin Lv** received the B.S., M.S., and Ph.D. degrees in electronic engineering from the Beijing Institute of Technology (BIT), in 1982, 1988, and 1993, respectively. Since 1982, he has been with BIT, as a Lecturer, an Associate Professor, and a Professor, and now he is retired.